\documentclass[11pt,epsf,letterpaper]{article}%
\usepackage{setspace}
\usepackage{color}
\usepackage{graphicx}
\usepackage{amsmath}
\usepackage{amsfonts}
\usepackage{verbatim}
\usepackage{amssymb}
\usepackage[normalem]{ulem}

\usepackage{url}

\usepackage[title]{appendix}

\providecommand{\U}[1]{\protect\rule{.1in}{.1in}}
\newcommand{\be}{\begin{equation}}
\newcommand{\ee}{\end{equation}}
\newcommand{\ba}{\begin{eqnarray}}
\newcommand{\ea}{\end{eqnarray}}

\begin{document}

\title{{\bf Feynman's {\it different} approach to electromagnetism}}

\author{Roberto De Luca$^{1}$, Marco Di Mauro$^{1,2,3}$,
Salvatore Esposito$^{2}$, \\
Adele Naddeo$^{2}$ \\
$^{1}$\footnotesize{Dipartimento di Fisica ``E.R. Caianiello",
Universit\'a di Salerno,
Via Giovanni Paolo II, 84084 Fisciano, Italy.}\\
$^{2}$\footnotesize{INFN Sezione di Napoli, Via Cinthia, 80126
Naples, Italy.}\\
$^{3}$E-mail: madimauro@unisa.it} \maketitle

\begin{abstract}
\noindent We discuss a previously unpublished description of
electromagnetism {outlined} by Richard P. Feynman in the 1960s in
five handwritten pages, recently uncovered among his papers, and
partly developed in later lectures. Though similar to the existing
approaches deriving electromagnetism from special relativity, the
present one extends a long way towards the derivation of Maxwell's
equations with minimal physical assumptions. {In particular,
without postulating Coulomb's law, homogeneous Maxwell's equations
are written down} by following a route different from the standard
one, i.e. first introducing electromagnetic potentials in order to
write down a relativistic invariant action, which is just the
inverse approach to the usual one. Also, Feynman's derivation of
the Lorentz force exclusively follows from its linearity in the
charge velocity and from relativistic invariance. {Going further,
i.e. adding the inhomogeneous Maxwell's equations, requires some
more physical input, and can be done by just following
conventional lines, hence this task was not pursued here. Despite
its incompleteness, this way of proceeding is of great historical
and epistemological significance. We also comment about its
possible relevance to didactics, as an interesting supplement to}
usual treatments.
\end{abstract}

\begin{flushright}
\emph{One man's assumption is another man's conclusion}
\cite{FeynmanHughes}
\end{flushright}

\section{Introduction}

It is a well-known story that Richard P. Feynman was not satisfied
with his presentation of electricity and magnetism in his second
year general physics lectures, which he gave at Caltech in
1961-63. In his preface to the published version
\cite{Feynman:1963uxa} he wrote:
\begin{quotation}
\noindent In the second year I was not so satisfied. In the first
part of the course, dealing with electricity and magnetism,
\emph{I couldn't think of any really unique or different way of
doing it} -- or any way that would be particularly more exciting
than the usual way of presenting it. So I don't think I did very
much in the lectures on electricity and magnetism.
\end{quotation}
As a matter of fact, indeed, the way of presenting
electromagnetism in the second volume of Feynman's Lectures
\cite{Feynman:1963uxa} is quite similar to many  existing books on
the subject at various level -- {for example the golden
standard} \cite{Jackson:1998nia} -- where a
historical/experimental path starting with electrostatics is
followed, then moving on to magnetostatics, Faraday's law of
induction, Maxwell's equations, electromagnetic waves and,
finally, showing the consistency of Maxwell's equations with the
special theory of relativity (which is however introduced in the
first volume of Ref. \cite{Feynman:1963uxa}).

Later on, in 1966, in an interview for the American Institute of
Physics with Charles Weiner \cite{FeynmanInterview1966},\footnote{The
relevant part of that interview is also reported in \cite{FeynmanTips}.}
Feynman said:
\begin{quote}
Now I think I know how to do it. [...] I've
now cooked up a much better way of presenting the electrodynamics,
a much more original and much more powerful way than is in the
book.
\end{quote}
Unfortunately, however, no such
presentation appears to have been published or used by Feynman.

Motivated by the natural curiosity aroused by such claims, M.A.
Gottlieb sought for an answer in the Caltech archives, where he
discovered five pages of handwritten notes, dating from 13
December 1963, in which Feynman sketched his ideas. {Since
it was natural for Gottlieb to assume that these notes are what Feynman
was referring to, without hesitation we here assume this to be the case.}
The notes (both the originals and a transcript by Gottlieb himself) were made
available online \cite{Gottlieb}. To the best of our knowledge, no
attempt to explain the physics contained in those notes was made
by Gottlieb or anyone else up to now, and thus, the aim of the
present paper is to study and clarify these notes, in order to
make their content usable. We believe that, besides the obvious
historical interest, this work can be useful for teaching
purposes, since it suggests an alternative view on at least part
of the fundamentals of electromagnetism.

Feynman's notes contain an outline of a possible course on
electromagnetism, along with some reflections on possible
disadvantages and advantages (interestingly, the disadvantages
come first), and a sketch of a derivation of the Lorentz force law
from the requirements of special relativity and of charge
invariance, which is the first point in the outline.

The idea of deriving electromagnetism from relativity (rather than
following the inverse, historical route) is not new, dating back
at least to 1912 \cite{Page}. The idea is to start from Coulomb's
law and then derive Lorentz force from Lorentz transformations.
While mentioned in most textbooks about electromagnetism, this
viewpoint has been chosen as the \emph{true} way to proceed in
many other textbooks and papers, see e.g. \cite{Page1940}-\cite{Susskind3}
and references therein.\footnote{Such an approach, in
order to be logically consistent, would of course require the
introduction of special relativity independently of
electromagnetism, which is indeed possible. For example, it was
used by Feynman himself in the first volume of his lectures, or in
many of the references cited in the main text. Moreover, it was
explored in many other references about special relativity, such
as Ref. \cite{Ugarov}, where special relativity is developed from
the idea that any interaction should have a limiting speed, only
later to be identified with the speed of light, and Ref.
\cite{Mermin}, where the relativistic law of addition of parallel
velocities is derived only from the first of Einstein's
postulates, while not invoking the second one.} Some of these
references content themselves with the derivation of
magnetostatics from electrostatic and relativity, while others try
to get the full electrodynamics by extending to accelerated
charges. Very interestingly, this same viewpoint has been adopted
by Feynman in teaching Electromagnetism at the Hughes Aircraft
company in 1967-68 \cite{FeynmanHughes}, and some comment about it
is also included in the chapter about magnetostatics in the second
volume of the lectures.

While sharing with the above viewpoint the idea of deriving
electromagnetism from relativity (rather than the opposite),
Feynman's approach sketched in his 1963 notes is radically
different. Feynman, indeed, derives the form of the interaction
(i.e.  Lorentz force) just from the requirements of charge
invariance (coming from experiments), relativity and linearity.
Such a derivation, to the best of our knowledge, is absent in the
literature, and was not even used by Feynman in later courses,
including the Hughes lectures. This was probably due to the fact
that he did not continue along this line of reasoning in the
derivation of Maxwell's equations and, more in general, of the
full electrodynamics. Nevertheless, some material in the Hughes
Lectures, as we shall see, might be in line with Feynman's ideas
in these notes, since, after a long introduction about the least
action principle (notoriously one of Feynman's favorites), he
described a relativistically invariant generalization of it where
force is implemented by the addition of a 4-vector potential. Upon
comparison of the resulting equations of motion with the Lorentz
force -- which may well have been derived using the approach
sketched in the notes, Feynman was able to find the link between
the electric and magnetic fields and the components of the
4-potential. These expressions, upon using well known vector
analysis identities, led to the homogeneous Maxwell equations.
Moreover, by allowing different potentials while requiring the
invariance of the classical equations of motions resulting from
minimizing the action, Feynman was able to introduce gauge
transformations from perspective different than the usual one. By
combining Feynman's approach to the Lorentz force with his
derivation of the homogeneous Maxwell's equations, it is then
possible to go quite a long way towards the full derivation of
electromagnetism, just with a minimal physical
input.\footnote{Interesting enough, this was not the only time
Feynman entertained himself with original presentations of
electromagnetism. As reported by Freeman Dyson (quoted in
\cite{DysonPhysTod}), in 1948 he was able to derive the
homogeneous Maxwell's equations solely from Newton's law and from
the canonical Poisson brackets (or even canonical commutation
relations) for a non-relativistic particle; such a derivation was
much later published by Dyson himself \cite{Dyson}, and was soon
recognized as a particular case of inverse variational problem
(see e.g. \cite{Carinena:1995ra} and references therein). Notice
that Feynman's proof of 1948 is given in a Galilei invariant
context, while the proof we are studying here is given in a
Lorentz invariant context. However, there is no contradiction
since both proofs concern only the Lorentz force and the
homogeneous Maxwell equations, which are invariant under both
groups (the deep origin of this is that the homogeneous Maxwell
equations are topological in nature, since they do not depend on
the metric, see e.g. \cite{Misner:1974qy}).} {After that,
it is possible to complete the construction of the theory by
adding the inhomogeneous Maxwell's equations in a more standard
way.}

\section{Deriving the Lorentz force}

In the first page of his notes, Feynman outlined a possible full
course on electrodynamics, articulated in seven points as follows:
\begin{enumerate}
\item Get the Lorentz force equation from charge
conservation\footnote{{Actually, from what follows it is
clear that here Feynman really means relativistic invariance of
the charge, rather than charge conservation, as in the usual sense.}} and relativity, get
the potentials or at least the homogeneous Maxwell equations.

\item Discuss the field idea, discuss the qualitative properties
and shapes of the fields in some situations, differential
operators, motion of electrons on given fields, induction, etc.

\item Get the inhomogeneous Maxwell's equations, either by
arriving at the wave equation for the potentials from relativity
or from some other principle, or even through the experimental,
usual route (Coulomb, Amp\`{e}re, \dots).

\item Discuss the fields produced in several semi-static
circumstances, e.g. condensers, inductances, etc.

\item Discuss the field in wave situations, e.g. radiation,
waveguides, etc.

\item Energy and related problems, self-mass, Lienard-Wiechert
potential, etc.

\item Fields in matter: $\mathbf{D}$, $\mathbf{H}$, etc.

\end{enumerate}
Points 1 and 3 are, of course, the crucial ones, as pointed out by
Feynman himself, while the other ones only concern applications of
electrodynamics, rather than the formulation of electromagnetic
laws, so that it is very much plausible that he aimed to address
them in a {standard} way, similar to the one adopted in the
Caltech or Hughes Lectures. Remarkably, however, in his notes
Feynman only addressed the first part of point 1, relegating the
second part just to a question (posed but not answered, at least
not in the notes): ``Can you find other ways? -- Can you get
potentials in?". Point 3 is not addressed as well.

Feynman's approach to electromagnetism can be traced back to the
fundamental assumptions that matter has an atomic structure (a
well-known citation shows that he considered this to be the single
most important scientific fact, as stated at the very beginning of
\cite{Feynman:1963uxa}), {and electric current is due to the
motion of electric charges in matter.}

\subsection{Linear dependence on the velocity}

The starting point, as delineated in the fifth page of his notes,
is that motion does not alter charge, i.e. electric charge is
Lorentz invariant. The usual argument for this, followed by
Feynman too, went as follows. Atoms are electrically neutral, but
electrons in them move, even at very high speeds, this meaning
that the electron charge is not affected by such motion. This is
in contrast with what happens for mass-energy, which does depend
on the velocity. Here Feynman referred properly to relativistic
energy, which may be described in terms of a velocity-dependent
``relativistic mass", considered as a gravitational mass (i.e.
weight) rather than inertial mass, though the two coincide due to
the equivalence principle. For atoms, this is exemplified by the
fact that an excited atom weights more than an atom in its ground
state, and Feynman also referred to a possible experimental proof
of this fact, while not explicitly mentioning it.\footnote{It is
worth noting that this same line of reasoning was followed by
Feynman in \cite{Feynman:1996kb}, in order to argue that
gravitation must be mediated by a spin-2 field rather than a
spin-0 field, whose charge would decrease with velocity, instead
of increasing.}

Having established Lorentz invariance for the electric charge, the
next step was the observation that a conducting wire, where  an
electric current is present, is electrically neutral. If two such
wires are placed next to each other, however, a net force between
them is observed (and here probably Feynman planned a
demonstration of this effect): this suggests that the force
between electric charges depends not only on their position, but
also on their velocity. Therefore, we conclude that a point charge
$q$, moving anyhow in the presence of other charges, experiences a
force $\mathbf{F}$ which is a function of its position and
velocity:
\begin{eqnarray}
\mathbf{F}=\mathbf{F}(\mathbf{x},\mathbf{v}) \, .
\end{eqnarray}
Charge invariance can be used again to observe that,
%\sout{for anoscillating charge, i.e.}
a charge moving in such a way that its
average velocity is zero {(such as for example an electron
in an atom), will feel an average force
$\overline{\mathbf{F}(\mathbf{x},\mathbf{v})}$ acting on it
because of the presence of other charges which must be
equal to the force felt by a charge of equal
magnitude at rest,\footnote{{Notice that we are considering
a charge moving in the presence of other charges, but the motion
is \emph{not} considered to be the consequence of the force
exerted by these other charges, rather of some other unspecified
external force.}}} i.e.
\begin{eqnarray}\label{condition}
\overline{\mathbf{F}(\mathbf{x},\mathbf{v})}=\mathbf{F}(\mathbf{\mathbf{x},0})
\qquad \text{if}\qquad \overline{\mathbf{v}}=0 \, .
\end{eqnarray}
This fact excludes the dependence of the force from any even power
of the velocity\footnote{As an example, let us show that for a
quadratic force the condition (\ref{condition}) is easily
violated. Consider a force of the form $F(v)=kqv^2$, where $k$ is
a constant and, as said, the charge $q$ does not depend on the
velocity . Suppose that the charge undergoes a periodic motion,
for example $v=v_0\cos(2\pi t/T)$. Then we clearly have
$\overline{v}=0$ over a period, but $ \overline{v^2}\neq 0$, hence
$\overline{F}\neq 0$, which contradicts the condition
(\ref{condition}). In a similar way all even powers of $v$ can be
excluded.}. Moreover, since this must be true independently of the
distribution of $\mathbf{v}$, provided only that the condition
$\overline{\mathbf{v}}=0$ holds, but without requirements on
higher moments, it necessarily implies that force depends {\it
linearly} on velocity,\footnote{The fact that the Lorentz force is
linear in the velocity was also emphasized by Feynman in chapter
12 of Volume I of his Lectures \cite{Feynman:1963uxa}, although
there he then derived the complete expression from experiment.}
that is $F_i(\mathbf{x},\mathbf{v})= a_i + b_{ij} v_j$, where
$a_i, b_{ij}$ are coefficients which can be cast in the more
expressive following form (making explicit the electric charge
$q$):
\begin{eqnarray}\label{force}
F_i(\mathbf{x},\mathbf{v})=q(E_i+v_j\,B_{ij}), \qquad i,j =1,2,3.
\end{eqnarray}
$E_i$ and $B_{ij}$ are just coefficients depending on the other
surrounding charges and on position, while the proportionality to
$q$ can be verified experimentally. {It is worth commenting
here that this line of reasoning seems to exclude damping terms
which may depend on higher powers of $\mathbf{v}$, but with
coefficients which go to zero very rapidly, i.e. with times much
smaller than the time scales considered above. Since the above
reasoning should be valid at least down to atomic scales, such
damping terms would be anyway out of the range of validity of
classical electromagnetism.}

{In hindsight, it is immediate to infer that $E_i$ are the
components of the electric field (or in other words, we can
call electric the part of the force which does not depend on the
velocity}), but the identification of $B_{ij}$ with the magnetic
induction (which is of course suggested by the notation) is not so
obvious: as a matter of fact, it constitutes the main part of
Feynman's notes. Crucial for this passage is to bring relativity
into play: the 3-vector in Eq. (\ref{force}) must be related to
the spatial components of a 4-force. By imposing that the Lorentz
transformation under a boost of this 4-force is linear in the
velocity, Feynman was able both to fix the functional form of the
force in (\ref{force}), thus recovering the Lorentz force, and to
derive the behavior of the coefficients $E_i$ and $B_{ij}$ under
the boost. And, not surprisingly, such transformation laws turn
out to be the correct ones for the electric and magnetic fields.

The reasoning can be sketched as follows. For simplicity we set
$q=1$; it is only a pre-factor that can be reinserted in any
moment (i.e. $\mathbf{F} \rightarrow q \mathbf{F}$). Also, we set
$c=1$, since it will always be possible to reinsert it by
dimensional analysis.

The 4-force associated with the 3-force in (\ref{force}) is
constructed using the well-known formulae from relativistic
mechanics, that is:
\begin{eqnarray}\label{4-force}
F^{\mu}= \gamma \left(F_t, F_x, F_y, F_z \right)  ,
\end{eqnarray}
where $F_t = \mathbf{F}\cdot{\mathbf{v}}$ and $\gamma=
1/\sqrt{1-v^2}$. Let us perform a boost with velocity
$\mathbf{u}$, for simplicity assumed directed along the $x$-axis:
\begin{equation}
\left\{
\begin{array}{rcl}
x' &=& \gamma_u (x-ut) , \\
y' &=& y , \\
z' &=& z , \\
t' &=& \gamma_u (t-ux) ,
\end{array} \right.
\qquad \qquad \left\{
\begin{array}{rcl}
v_x' &=& \displaystyle \frac{v_x-u}{1-uv_x} \, , \\ & & \\
v_y' &=& \displaystyle \frac{v_y}{\gamma_u(1-uv_x)}\, ,\\ & & \\
v_z' &=& \displaystyle \frac{v_z}{\gamma_u(1-uv_x)} \, ,
\end{array} \right.
\label{Lorentz1}
\end{equation}
where $\gamma_u=1/\sqrt{1-u^2}$ is the gamma factor associated
with the boost velocity $\mathbf{u}$. Under such a boost, the
component $x$ of the 4-vector (\ref{4-force}) transforms in the
usual way, that is
%$\gamma' F'_x = \gamma_u \gamma \left( F_x - u F_t \right)$, where $\gamma' = 1/\sqrt{1-v^{' 2}}$, or
%
\begin{eqnarray}\label{Fprimex_0}
F'_x=\frac{F_x-uF_t}{1-uv_x} \, .
\end{eqnarray}
By inserting the explicit expressions for $F_x$ and $F_t$, and
expressing the result in terms of only the primed velocity
components, after some algebra\footnote{\label{foot1}Useful
relations, which may be employed in order to reduce calculations,
are the following:
\[
1 - u v_x = \frac{\gamma'}{\gamma \gamma_u} \, , \qquad (1 + u
v_x) (1 - u v_x) = 1 - u^2  ,
\]
where $\gamma' = 1/\sqrt{1-v^{' 2}}$.} we arrive at:
\begin{eqnarray}\label{Fprimex}
F'_x&=&E_x-u\gamma_uv'_yE_y-u\gamma_u
v'_zE_z+\gamma_uv'_yB_{xy}+\gamma_uv'_zB_{xz} +
\frac{v'_x+u}{1+uv'_x}B_{xx}-\frac{u(v'_y)^2B_{yy}}{1+uv'_x}-\frac{u(v_z')^2B_{zz}}{1+uv'_x}
\nonumber \\ &&-u\gamma_u
\frac{(v'_x+u)v'_y}{1+uv'_x}(B_{xy}+B_{yx})
-\frac{uv'_yv'_z}{1+uv'_x}(B_{yz}+B_{zy})-u\gamma_u\frac{(v'_x+u)v'_z}{1+uv'_x}(B_{xz}+B_{zx}).
\end{eqnarray}
We have now to require that $F_x'$ must have the same form as
$F_x$, that is
\begin{eqnarray}\label{Fprimex2}
F_x'=E'_x+v'_xB'_{xx}+v'_yB'_{xy}+v'_zB'_{xz} \, .
\end{eqnarray}
Since the argument requiring linearity in the velocity should be
valid in any inertial frame, this implies that all the terms of
(\ref{Fprimex}) that are quadratic in the velocity components
should vanish; the same applies to the term in (\ref{Fprimex})
with no component of $\mathbf{v}'$ in the numerator. Then, the $B$
coefficients have to satisfy the following relations:
\begin{equation}\label{cond123}
\begin{array}{c}
B_{xy}+B_{yx}=0,\qquad B_{yz}+B_{zy}=0, \qquad B_{xz}+B_{zx}=0 , \\
B_{xx}=0, \qquad B_{yy}=0, \qquad B_{zz}=0.
\end{array}
\end{equation}
Such conditions (\ref{cond123}) imply that the $B$ coefficients
are antisymmetric, i.e. $B_{ij}=-B_{ji}$. It is customary to
rearrange the three independent components of an antisymmetric
matrix in the form of a $3$-component vector, defined in the
present case by $B_i=\frac{1}{2}\epsilon_{ijk}B_{jk}$, or $B_x =
B_{yz}$, $B_y = B_{zx}$, $B_z = B_{xy}$. This allows us to write
the force (\ref{force}) in the familiar form (where we restore the
$q$ dependence)
\begin{eqnarray}\label{CoulombLorentz}
\mathbf{F}=q \left( \mathbf{E}+\mathbf{v}\times\mathbf{B} \right)
,
\end{eqnarray}
which is the usual expression for the Lorentz force acting on the
charge $q$, {\it provided} we identify the coefficients $E$ and
$B$ with the electric and magnetic fields respectively.

\subsection{Relativistic invariance and the E, B fields}

Such identification requires that those quantities transform
correctly under Lorentz boosts, and Feynman's proof proceeds as
follows. First of all, by comparing the non-vanishing terms in
(\ref{Fprimex}) (after imposing conditions (\ref{cond123})) with
the analogous terms in (\ref{Fprimex2}), we can infer the behavior
of some of the $E$, $B$ coefficients under a boost, i.e.
\begin{eqnarray}\label{cond4}
E'_x=E_x, \qquad B'_y=\gamma_u(B_y+uE_z), \qquad
B'_z=\gamma_u(B_z-uE_y) .
\end{eqnarray}
We recognize here just the usual transformation laws of some of
the components of the electric and magnetic field under a boost
with velocity $\mathbf{u}$ directed along the positive $x$-axis:
in the present framework, they directly follow from the linearity
in the velocity and relativistic mechanics. However, we are not
done yet, since the transformation rules for $E_y$, $E_z$ and
$B_x$ still lack. The first two come from requiring that the
fourth component of the $4$-vector (\ref{4-force}) transform in
the usual way, i.e. $\gamma' F'_t=\gamma_u \gamma (F_t -u F_x)$.
By exploiting the conditions (\ref{cond123}) and (\ref{cond4}) and
the Lorentz transformation laws of the fields we have just
derived, as well as the fact that, because of
(\ref{CoulombLorentz}), $F_t=\mathbf{E}\cdot\mathbf{v}$ ($q=1$),
along the same lines as above we obtain:
\begin{eqnarray}
F'_t&=&\frac{1}{1-u
v_x}[E_x(v_x-u)+E_yv_y+E_zv_z-uv_yB_z+uv_zB_y]\nonumber\\&=&
v'_xE'_x + v'_y\gamma_u(E_y-uB_z)+v'_z\gamma_u(E_z+uB_y).
\end{eqnarray}
And, by imposing that
$F'_t=\mathbf{F}'\cdot\mathbf{v}'=\mathbf{E}'\cdot\mathbf{v}'$,
the following transformation rules follow
\begin{eqnarray}
E'_y= \gamma_u(E_y-uB_z), \qquad E'_z=\gamma_u(E_z+uB_y),
\end{eqnarray}
which again are the expected ones. The last transformation rule
for $B_x$ can be obtained by considering the third or fourth
component of the $4$-vector (\ref{4-force}). For the boost
considered here, such components must be invariant, they being
orthogonal to the boost velocity, so that $\gamma' F'_y = \gamma
F_y$. By proceeding as above, we get
\begin{eqnarray}
F'_y&=&\frac{1}{\gamma_u
(1-uv_x)}(E_y-v_xB_z+v_zB_x)\nonumber\\&=&
\gamma_u(1+uv'_x)E_y-\gamma_u(v'_x+u)B_z+v'_zB_x ,
\end{eqnarray}
or, by using the inverse transformation laws of $E_y$ and
$B_z$,\footnote{That is: $E_y=\gamma_u(E'_y+uB'_z)$,
$B_z=\gamma_u(B'_z+uE'_y)$.}
\begin{eqnarray}
F'_y=[\gamma^2_u(1+uv'_x)-\gamma_u^2 u v'_x-\gamma_u
u^2]E'_y+[\gamma_u^2(1+uv_x')u-\gamma_u^2v'_x-\gamma_u^2u]B'_z+v'_zB_x
.
\end{eqnarray}
Upon simple manipulation, this reduces to
$F'_y=E'_y-v'_xB'_z+v'_zB_x$, so that by imposing this to be of
the form $F'_y=E'_y-v'_xB'_z+v'_zB'_x$, we finally get the last
transformation law,
\begin{eqnarray}
B'_x=B_x,
\end{eqnarray}
as expected.

\section{Least action principle}

A possible way to address the second part of Feynman's point 1
above -- i.e. introduce the homogeneous Maxwell equations --can be
traced in his Hughes lectures \cite{FeynmanHughes}, where the
vector potential was there introduced in order to write down a
relativistically invariant least action principle for a particle
in a given potential {energy $\cal V$}. It is conceivable that
Feynman had something like this in mind in 1963, when he wrote his
notes, but it is as well possible that he thought about this way
of proceeding sometime between 1963 and 1967. Part of this
discussion is already present in Chapter 19 of Volume II of the
Lectures \cite{Feynman:1963uxa}, though in a much less detail;
similar reasoning is also taken up e.g. by Susskind in his
Theoretical Minimum \cite{Susskind3}. {As is well known,
Feynman was convinced that potentials had the same level of
reality as the fields. Quite illuminating, in this respect, is what quoted in \cite{Goodstein}:
\begin{quote}
Yet, the Schr\"odinger equation can only be written neatly with $\bf A$ and $V$ explicitly there
and it was pointed out by Bohm and Aharonov (or something like that), that this means that the vector
potential has a reality and that in quantum mechanical interference experiments there can be
situations in which classically there would be no expected influence whatever. But nevertheless
there is an influence. Is it action at a distance? No, $\bf A$ is as real as $\bf B$-realer, whatever that means.
\end{quote}
}

\subsection{Homogeneous Maxwell equations}

The starting point is the generalization of the least action
principle in classical mechanics to the relativistic case. Since
in relativistic mechanics the momentum of a particle of mass $m_0$
is given by ${m_0\mathbf{\dot{x}}}/{\sqrt{1-v^2}}$, as a first
naive guess we may seek for an action $S$ from which the equations
of motion
\begin{eqnarray}\label{RelEOM1}
\frac{\rm d}{{\rm
d}t}\left[\frac{m_0\mathbf{v}}{\sqrt{1-v^2}}\right]=-\nabla {\cal V},
\end{eqnarray}
can be obtained. Such an action may be identified as
\begin{eqnarray}\label{RelAct1}
S=\int_{t_1}^{t_2}\left[-m_0\sqrt{1-v^2}-{\cal V}(\mathbf{x},t)\right]
{\rm d}t = \int_{t_1}^{t_2} [-m_0 \, {\rm d}s - {\cal V}(\mathbf{x},t) \,
{\rm d}t] \,
\end{eqnarray}
(${\rm d}s=\sqrt{1-v^2} \, {\rm d}t$), since it can be easily
checked that, indeed, Eq. (\ref{RelEOM1}) just follows from
requiring the expression in (\ref{RelAct1}) to be an extremum. The
point here, however, is the relativistic invariance of the action
or, in Feynman's words, to establish if the action keeps its
extremum in any inertial reference frame. Since the potential
{energy} term is evidently non invariant, it must be
modified. {The obvious choice would be an invariant scalar
potential term $\mathcal{X}(x,y,z,t) \, {\rm d}s$  but, as
quickly stated by Feynman, such a term does not lead to any known
law of nature. A more elaborate discussion on scalar field
forces can be found in \cite{Feynman:1996kb} and
\cite{FeynmanHughes1}, where Feynman notices that such fields
would require sources which -- unlike the electric charge --
decrease with the velocity. As} the next simplest possibility,
Feynman then suggested the use of a $4$-potential
$A_{\mu}(x,y,z,t)$, in order for the action to assume the
form\footnote{Feynman briefly goes on suggesting a $10$-potential
term of the form $-g_{tt}\frac{{\rm d}t}{{\rm d}s}\frac{{\rm
d}t}{{\rm d}s}{\rm d}s-g_{tx}\frac{{\rm d}t}{d{\rm s}}\frac{{\rm
d}x}{{\rm d}s}{\rm d}s-\ldots$, mentioning that such term allows
to derive the force of gravity. It is interesting to note that in
\cite{FeynmanHughes1} Feynman also gives some references to an
approach to gravity which mimics the one given here for
electrodynamics, in particular suggesting that in the case of
gravity Eq. (\ref{force}) should be replaced by a quadratic
expression.}
\begin{eqnarray}\label{RelAct2}
S = \int_{t_1}^{t_2} [-m_0 \, {\rm d}s - A_t \, {\rm d}t+A_x \,
{\rm d}x+A_y \, {\rm d}y+A_z \,{\rm d}z] = \int_{t_1}^{t_2} [-m_0
\, {\rm d}s - A_{\mu} \, {\rm d}x^{\mu}] \, .
\end{eqnarray}
In the following we shall set $A_t=\phi$, thus matching the usual
notation. The next step is, of course, to vary this action in
order to see what equations of motion result; this is done in
Appendix \ref{AppC}, and the result is the equation
\begin{eqnarray}\label{RelEOM2}
\frac{\rm d}{{\rm
d}t}\left[\frac{m_0\mathbf{v}}{\sqrt{1-v^2}}\right]= \mathbf{F},
\end{eqnarray}
with the force $\mathbf{F}$ given by:
\begin{eqnarray}\label{EoM}
\mathbf{F}=-\nabla\phi-\frac{\partial}{\partial t}\mathbf{A} +
\mathbf{v}\times(\nabla \times \mathbf{A}).
\end{eqnarray}
Such expression looks just like that of the Lorentz force
(provided we restore $q$ in front of it, amounting to use
$qA_{\mu}{\rm d}x^{\mu}$ in place of $A_{\mu}{\rm d}x^{\mu}$ in
the action), when we identify:
\begin{eqnarray}\label{fields}
\mathbf{E}=-\nabla\phi-\frac{\partial}{\partial
t}\mathbf{A};\qquad \mathbf{B}=\nabla\times\mathbf{A} \, .
\end{eqnarray}
Using well-known vector analysis formulae, namely
$\nabla\cdot(\nabla \times \mathbf{V})=0$ and $\nabla\times(\nabla
f)=0$, valid for any vector field $\mathbf{V}$ and any scalar
function $f$, we can then write down the homogeneous Maxwell
equations:
\begin{eqnarray}\label{HomogeneousMaxwell}
\nabla\cdot\mathbf{B}=0,\qquad \nabla\times\mathbf{E}=
-\frac{\partial}{\partial t}\mathbf{B}.
\end{eqnarray}

\subsection{Gauge transformations}

In the Hughes lectures, Feynman deepened his discussion of the
least action principle by discussing the gauge transformations of
the potentials. Starting again from the action (\ref{RelAct2}),
which is now rewritten as
\begin{eqnarray}
S=-m_0\int \! {\rm d}s + \int \! (\phi  -
\mathbf{A}\cdot\mathbf{v})\, {\rm d}t \, ,
\end{eqnarray}
we can ask whether we could use a different set of potentials
$\phi'$, $\mathbf{A}'$ and still get the same physical trajectory,
i.e. the same minimum of the action. This evidently happens if the
difference of the action written in terms of the first set of
potentials and the action written in terms of the second set of
potentials is independent of the path. Such difference,
\begin{eqnarray}\label{difference1}
S-S'=\int \! (\varphi -\mathbf{a}\cdot\mathbf{v}) \, {\rm d}t \, ,
\end{eqnarray}
($\varphi=\phi-\phi'$, $\mathbf{a}=\mathbf{A}-\mathbf{A}'$) is
independent of the integration path if the integrand is a perfect
differential, i.e. $\varphi -\mathbf{a}\cdot\mathbf{v}={{\rm
d}\chi(x,y,z,t)}/{{\rm d}t}$, so that
\begin{eqnarray}\label{difference2}
S-S'&=&\int_{t_i}^{t_f} \frac{{\rm d}\chi(x,y,z,t)}{{\rm
d}t}=\int_{t_i}^{t_f} \left(\frac{\partial\chi}{\partial t} +
\frac{\partial\chi}{\partial x}\dot{x} +
\frac{\partial\chi}{\partial y}\dot{y}+
\frac{\partial\chi}{\partial z}\dot{z}\right) {\rm d}t \nonumber \\
&=& \chi(x,y,z,t_f)-\chi(x,y,z,t_i) \, .
\end{eqnarray}
Comparing (\ref{difference1}) with (\ref{difference2}) we see that
such condition is satisfied if $\varphi ={\partial \chi}/{\partial
t}$ and $\mathbf{a}=-\nabla \chi$, so that the two actions $S$ and
$S'$ lead to the same equations of motion if the potentials are
related by the gauge transformations
\begin{eqnarray}
\phi'=\phi + \frac{\partial \chi}{\partial t}\, , \qquad
\mathbf{A}'=\mathbf{A}-\nabla \chi \, .
\end{eqnarray}
The fact that the equations of motion do not change under such
transformations is also evident from the fact that the electric
and magnetic fields defined in (\ref{fields}), which are just the
quantities appearing in the equations of motion, are left
unchanged. Quite noticeable is this reversed line of reasoning
with respect to the standard presentation; {the latter} was
indeed followed by Feynman, in particular, in his Caltech
lectures.

\section{Discussion and conclusions}

We have discussed {and partially developed} a previously
unpublished description of electromagnetism {outlined} by Richard
Feynman. {Notwithstanding the fact that such description was left
incomplete, the key epistemological interest in the exploration of
alternative formulations of classical electrodynamics is evident
by itself. How much can we deduce from charge conservation alone,
or from special relativity, or from the more elusive assumptions
of Lagrangian field theory? Much genuine physical insight can be
derived from such questions.}

Motivated by his involvement in undergraduate physics teaching at
Caltech in 1961-63, Feynman obtained what we described above only
in late 1963, when this involvement was already over, so that it
was definitively not published. The approach adopted was similar
to the existing ones deriving electromagnetism from special
relativity, but manages to go a long way towards {obtaining}
Maxwell's equations with minimal physical assumptions, in
particular without postulating Coulomb's law. {We do not know why
Feynman thought Coulomb's law was -- in a sense -- so wicked, but
we may argue that his foundation of electrodynamics upon forces
between two (electrically neutral) current-carrying wires, instead
of forces between two charged particles, was favored by the
certainly simpler {\it experimental} definition of currents
compared to that of pointlike charges, irrespective of his simpler
{\it conceptual} atomic vision of electric currents as due to the
motion of electric charges in matter, which evidently is more
abstract in nature.} To the best of our knowledge, after sketching
such an approach, Feynman never used it, even when he taught
electromagnetism again, at the Hughes Aircraft Company in 1967. In
the latter lectures, however, he adopted an approach which was
closer in spirit to the one described here.

As evident from what reported in the previous sections, Feynman's
description of electromagnetism was left incomplete, the part
remaining in classical electrodynamics discussing the
inhomogeneous Maxwell equations: this would require to go well
beyond what has been done so far. Indeed, electric and magnetic
fields have been considered merely with respect to their effects
on charged particles, this allowing to go quite far, up to the
homogeneous Maxwell equations and the related gauge
transformations, but this is only half of the story. As a matter
of fact, it is as well necessary to consider the effect of charges
on the fields, i.e. what kind of electric and magnetic fields are
generated by charges, and this is evidently the content of
inhomogeneous Maxwell's equations. Feynman was of course aware
that some new input was necessary at this stage, so that in point
$3$ of the outline above he proposed to follow the usual
experimental route going through Coulomb's inverse square law and
Amp\`{e}re's law for the force between currents. He also suggested
that it may be possible to arrive at the wave equation for the
potentials (but how?). {We do not know whether Feynman
thought about this point, but it would certainly be of interest to
try to develop it, even though we may never know if and how the
result will be close to Feynman's ideas}). {\it De facto}, in the
derivation above we have not used any experimental input about the
electric and magnetic forces, except the fact that they exist and
depend on charge and velocity. In all attempts in the literature
to deduce magnetostatics or even full electrodynamics from
relativity, one always starts from the experimental Coulomb law,
{or at least, as in \cite{Landau:1982dva}, from the
additional requirement (which again comes from experiment) that
the superposition principle holds, which, together with
covariance, forces the action \emph{for the electromagnetic field}
to be linear.}

Bringing to light this approach is of obvious historical
significance, adding a further piece to the already variegated
Feynman jigsaw. However, in view of its originality, it is not
inconceivable that this work can have some interest from the point
of view of physics teaching. While this approach is clearly not
suitable for a first course, it provides a novel way to develop
part of the foundations of electromagnetism, which might
fruitfully supplement a more conventional treatment. The most
original points are, in our view, the derivation of the Lorentz
force and the inverse route in the derivation of the homogeneous
Maxwell's equation, i.e. first introducing electromagnetic
potentials in order to write down a relativistically invariant
action. A very interesting development of this work would then
involve putting these ideas to test in the classroom. For example,
some bits of an advanced undergraduate course in electromagnetism
and special relativity could be devoted to an exposition of the
matter following the lines discussed in the present paper, and
monitoring the response of the students to it.

It is certainly highly desirable that other pieces of Feynman
``magic" come to light in the near future, {especially concerning
the completion of this picture}. In particular, from
\cite{FeynmanHughes}, \cite{Feynman:1996kb}, and
\cite{FeynmanHughes1}, as remarked above, we found evidence that
an approach to gravity which is analogous to the one discussed
here for electrodynamics was in Feynman's thoughts. The
development of such an approach will be the subject of a
forthcoming publication \cite{Noi}.

\begin{appendices}

\numberwithin{equation}{section}
\section{Useful formulae}

Consider a boost with velocity $\mathbf{u}$ in the positive $x-$
direction
\begin{eqnarray}
x'&=&\gamma_u(x-ut)\\
y'&=&y\nonumber\\
z'&=&z\nonumber\\
t'&=&\gamma_u(t-ux)\nonumber
\end{eqnarray}
where $\gamma_u=(1-u^2)^{-1/2}$ is the gamma factor associated
with the boost velocity $\mathbf{u}$. Then the components of the
velocity $\mathbf{v}$ of a particle transform in the well known
way:
\begin{eqnarray}\label{Lorentz1}
v_x'&=&\frac{v_x-u}{1-uv_x}, \\
v_y'&=&\frac{v_y}{\gamma_u(1-uv_x)},\nonumber\\
v_z'&=&\frac{v_z}{\gamma_u(1-uv_x)}.\nonumber
\end{eqnarray}
Using these relations, it is straightforward to prove the
following relations:
\begin{eqnarray}\label{Lorentz2}
1+uv_x'=1+u\frac{v_x-u}{1-uv_x}=\frac{1-u^2}{1-uv_x}
\end{eqnarray}
\begin{eqnarray}\label{useful}
\sqrt{1-(v')^2}=\sqrt{1-(v_x')^2-(v_y')^2-(v_z')^2}=\frac{\sqrt{1-v^2}\,\sqrt{1-u^2}}{1-uv_x}
\end{eqnarray}
These relations will be useful in the calculations performed in
the main text. For convenience, we also write down the inverses of
(\ref{Lorentz1}) and (\ref{Lorentz2}), which can be obtained
simply by exchanging primed and unprimed components of
$\mathbf{v}$ and inverting $u\rightarrow -u$:
\begin{eqnarray}\label{Lorentz1inv}
v_x&=&\frac{v'_x+u}{1+uv'_x}, \\
v_y&=&\frac{v'_y}{\gamma_u(1+uv'_x)},\nonumber\\
v_z&=&\frac{v'_z}{\gamma_u(1+uv'_x)},\nonumber
\end{eqnarray}
\begin{eqnarray}\label{Lorentz2inv}
1-uv_x=1-u\frac{v'_x+u}{1+uv'_x}=\frac{1-u^2}{1+uv'_x} \qquad
\text{or} \qquad \frac{1}{1-uv_x}=\gamma_u^2(1+uv'_x).
\end{eqnarray}

\section{Some details on the derivation of (\ref{Fprimex})}

The explicit expressions for $F_x$, $F_y$, $F_z$ and $F_t$ are
respectively (we put $q=1$):
\begin{eqnarray}
F_x&=&E_x+v_xB_{xx}+v_yB_{xy} + v_zB_{xz}\\
F_y&=&E_y+v_xB_{yx}+v_yB_{yy} + v_zB_{yz}\\
F_z&=&E_z+v_xB_{zx}+v_yB_{zy} + v_zB_{zz}
\end{eqnarray}
and
\begin{eqnarray}
F_t&=&\mathbf{v}\cdot\mathbf{F}\nonumber\\&=&v_x(E_x+v_xB_{xx}+v_yB_{xy}
+ v_zB_{xz}) +v_y(E_y+v_xB_{yx}+v_yB_{yy} + v_zB_{yz})
\nonumber\\&&+v_z(E_z+v_xB_{zx}+v_yB_{zy} + v_zB_{zz}).
\end{eqnarray}
Therefore, substituting in (\ref{Fprimex_0}), we get
\begin{eqnarray}
F'_x&=&\frac{1}{1-uv_x}[E_x+v_xB_{xx}+v_yB_{xy} +
v_zB_{xz}-uv_x(E_x+v_xB_{xx}+v_yB_{xy} + v_zB_{xz})\nonumber\\
&&-uv_y(E_y+v_xB_{yx}+v_yB_{yy} + v_zB_{yz})
-uv_z(E_z+v_xB_{zx}+v_yB_{zy} + v_zB_{zz})]\nonumber\\ &=&
\frac{1}{1-uv_x}[E_x(1-uv_x)-uv_yE_y-uv_zE_z+v_x(1-uv_x)B_{xx}+v_yB_{xy}+v_zB_{xz}\nonumber\\&&-uv_xv_y(B_{xy}+B_{yx})-uv_zv_x(B_{xz}+B_{zx})-uv_y^2B_{yy}-uv_z^2B_{zz}-uv_xv_z(B_{yz}+B_{zy})].
\end{eqnarray}
Using now (\ref{Lorentz1inv}) and (\ref{Lorentz2inv}) we can put
everything in terms of the primed velocity components. After some
straightforward algebra we arrive at (\ref{Fprimex}).

\section{Variational calculations} \label{AppC}

By following \cite{FeynmanHughes}, we here find the Euler-Lagrange
equations for the action (\ref{RelAct2}), which we rewrite here as
follows:
\begin{eqnarray}
S = -\int_{t_i}^{t_f} \left(
-m_0\sqrt{1-\dot{x}^2-\dot{y}^2-\dot{z}^2}
-\phi+A_x\dot{x}+A_y\dot{y}+A_z\dot{z} \right) {\rm d}t
\end{eqnarray}
We have to vary the path $(x(t), y(t), z(t))$ by infinitesimal
quantities $(\xi(t), \eta(t), \zeta(t))$,
\begin{eqnarray}
x(t)\rightarrow x(t)+\xi(t),\quad y(t)\rightarrow
y(t)+\eta(t),\quad z(t)\rightarrow z(t)+\zeta(t),
\end{eqnarray}
with the constraint that the variations vanish at the endpoints
$t_i, t_f$, and keep only first order terms in $\xi$, $\eta$ and
$\zeta$ and their time derivatives. This induces the action $S$ to
vary as well by $S\rightarrow S + \delta S$ and, by requiring it
to be stationary, the equations of motion for all three
coordinates $x$, $y$ and $z$ follow. This procedure, however,
would involve a lot of terms, so that, by following Feynman, we
choose to vary only one coordinate at a time, in order to get only
one component of the equation of motion: since calculations are
very similar in the three cases, we here perform it only for the
$x$ case, leaving the other two to the reader. We have:
\begin{eqnarray}
S+\delta S&=&\int_{t_i}^{t_f} \left[-m_0
\sqrt{1-(\dot{x}+\dot{\xi})^2-\dot{y}^2-\dot{z}^2} \, - \, \phi(x+\xi,y,z,t) \, +\, A_x(x+\xi,y,z)(\dot{x}+\dot{\xi}) \right. \nonumber \\
&& \left. + \, A_y(x+\xi,y,z,t)\dot{y} \, + \,
A_z(x+\xi,y,z,t)\dot{z}
\vphantom{\sqrt{1-(\dot{x}+\dot{\xi})^2-\dot{y}^2-\dot{z}^2}}
\right] {\rm d}t.
\end{eqnarray}
By Taylor expanding at first order the integrand, we have:
\begin{eqnarray}
\sqrt{1-(\dot{x}+\dot{\xi})^2-\dot{y}^2-\dot{z}^2} & \simeq &
\sqrt{1-v^2-2\dot{x}\dot{\xi}} \, \simeq \, \sqrt{1-v^2} \left( 1
-\frac{\dot{x}\dot{\xi}}{1-v^2} \right) \, = \, \sqrt{1-v^2} -
\frac{\dot{x}\dot{\xi}}{\sqrt{1-v^2}} , \nonumber
\\
\phi(x+\xi,y,z,t) & \simeq & \phi(x,y,z,t) \, + \, \xi \,
\frac{\partial\phi}{\partial x}(x,y,z,t) , \nonumber
\\
A_x(x+\xi,y,z,t) & \simeq & A_x(x,y,z,t) \, + \, \xi \,
\frac{\partial A_x}{\partial x}(x,y,z,t) , \nonumber
\\
A_y(x+\xi,y,z,t) & \simeq & A_y(x,y,z,t) \, + \, \xi \,
\frac{\partial A_y}{\partial x}(x,y,z,t) , \nonumber
\\
A_z(x+\xi,y,z,t) & \simeq & A_z(x,y,z,t) \, + \, \xi \,
\frac{\partial A_z}{\partial x}(x,y,z,t) , \nonumber
\end{eqnarray}
so that:
\begin{eqnarray}
S+\delta S &\simeq& \int_{t_i}^{t_f} \left( -m_0 \sqrt{1-v^2} -\phi+A_x\dot{x}+A_y\dot{y}+A_z\dot{z} \right) {\rm d}t  \nonumber \\
&& + \,  \int_{t_i}^{t_f} \left(
\frac{m_0\dot{x}}{\sqrt{1-v^2}}+A_x\right)\dot{\xi}\,{\rm d}t \, +
\, \int_{t_i}^{t_f} \xi \left(-\frac{\partial \phi}{\partial x}+
\frac{\partial A_x}{\partial x}\dot{x}+\frac{\partial
A_y}{\partial x}\dot{y} + \frac{\partial A_z}{\partial
x}\dot{z}\right) {\rm d}t.
\end{eqnarray}
The first integral build up again the action $S$ we started with,
while the second one can be integrated by parts to give:
\begin{eqnarray}
\int_{t_i}^{t_f}\left(\frac{m_0\dot{x}}{\sqrt{1-v^2}}+A_x\right)\dot{\xi}\,
{\rm d}t=
\left[\left(\frac{m_0\dot{x}}{\sqrt{1-v^2}}+A_x\right)\xi\right]^{t_f}_{t_i}-\int_{t_i}^{t_f}\xi\frac{\rm
d}{{\rm d}t}\left(\frac{m_0\dot{x}}{\sqrt{1-v^2}}+A_x\right) {\rm
d}t,
\end{eqnarray}
where the first term vanishes since we assumed that at the
endpoints $\xi(t_i)=\xi(t_f)=0$. Therefore we get:
\begin{eqnarray}
\delta S=\int \xi\left(-\frac{\rm d}{{\rm
d}t}\frac{m_0\dot{x}}{\sqrt{1-v^2}}-\dot{A}_x-\frac{\partial
\phi}{\partial x}+ \frac{\partial A_x}{\partial x}\,
\dot{x}+\frac{\partial A_y}{\partial x} \, \dot{y} +
\frac{\partial A_z}{\partial x} \, \dot{z}\right)\, {\rm d}t.
\end{eqnarray}
Here, $\dot{A}_x$ is a total derivative, i.e.
\begin{eqnarray}
\dot{A}_x=\frac{\partial A_x}{\partial x}\dot{x}+ \frac{\partial
A_x}{\partial y}\dot{y}+ \frac{\partial A_x}{\partial z}\dot{z}
+\frac{\partial A_x}{\partial t} \, .
\end{eqnarray}
By requiring $\delta S=0$, {since the variation of $\xi$
is} arbitrary, the equation of motion follows:
\begin{eqnarray}
\frac{\rm d}{{\rm d}t}\left[\frac{m_0\dot{x}}{\sqrt{1-v^2}}\right]
&=&-\frac{\partial \phi}{\partial x}-\frac{\partial A_x}{\partial
t} + \dot{y}\left( \frac{\partial A_y}{\partial x}-\frac{\partial
A_x}{\partial y}\right) + \dot{z}\left( \frac{\partial
A_z}{\partial x}-\frac{\partial A_x}{\partial z} \right).
\end{eqnarray}
Similarly, by requiring the action to be stationary when varying
$y(t)$ and $z(t)$, we arrive at the  other two equations of
motion:
\begin{eqnarray}
\frac{\rm d}{{\rm
d}t}\left[\frac{m_0\dot{y}}{\sqrt{1-v^2}}\right]&=&-\frac{\partial
\phi}{\partial y}-\frac{\partial A_y}{\partial t}  + \dot{z}\left(
\frac{\partial A_z}{\partial y}-\frac{\partial A_y}{\partial z}
\right) + \dot{x}\left( \frac{\partial A_x}{\partial y} -
\frac{\partial A_y}{\partial x} \right) ,
\\
\frac{\rm d}{{\rm
d}t}\left[\frac{m_0\dot{z}}{\sqrt{1-v^2}}\right]&=&-\frac{\partial
\phi}{\partial z}-\frac{\partial A_z}{\partial t} + \dot{x}\left(
\frac{\partial A_x}{\partial z}-\frac{\partial A_z}{\partial
x}\right) + \dot{y}\left( \frac{\partial A_y}{\partial
z}-\frac{\partial A_z}{\partial y} \right).
\end{eqnarray}
In vector form, these three equations of motion rewrite as
\begin{eqnarray}
\frac{\rm d}{{\rm
d}t}\left[\frac{m_0\mathbf{v}}{\sqrt{1-v^2}}\right]=\mathbf{F}
\end{eqnarray}
with $\displaystyle
\mathbf{F}=-\nabla\phi-\frac{\partial}{\partial t}\mathbf{A} +
\mathbf{v}\times(\nabla \times \mathbf{A})$,  as in Eq.
(\ref{EoM}).

\end{appendices}

\end{document}